# Measuring magnetic field inside a microwave cavity via Rabi resonances in Cs atoms

Fuyu Sun*, Dong Hou, Qingsong Bai, and Xianhe Huang

*Time & Frequency Research Center, School of Automation Engineering, University of Electronic Science and Technology of China, Chengdu, 611731, China*

Jie Ma, and Xiaofeng Li

*Key Laboratory of Time and Frequency Primary Standards, National Time Service Center, Chinese Academy of Science, Xi'an, Shaanxi, 710600, China*

We present a technique for measuring microwave (MW) field based on Rabi resonances induced by the interaction of atoms with a phase-modulated MW field. A theoretical model of field measurement is used to calculate Rabi frequency. Single-peak feature of the measurement model makes the technique a valuable tool for simple and fast field measurement. As an example, we use the technique to determine the MW field strength inside a Cs vapor cell in the X-band rectangular cavity for applied power in the range of -21 dBm to 20 dBm. The results show that this proposed technique is capable for detecting the field over a broad dynamical range.

## I. INTRODUCTION

To date, atom-based measurements for static and radio frequency fields have already made remarkable successes [1-13]. Recently, atom-based microwave (MW) measurement has also inspired great interest because of its potential ability to link the MW quantities with SI second. As a result, relying on various physical principles, many atomic MW sensors have been developed, such as MW power standard [14-18], MW electrometry [19-27], MW electric/magnetic field imaging [21,28-33], and MW magnetometer [34]. As compared to traditional measurement, atom-based measurement is intrinsically calibrated where field strength is translated into Rabi frequency $\Omega$ via well-known atomic constants. Particularly, the use of atomic vapor cell, serves as sensing head, making atom-based measurements properly for practical applications [19,24,25,29]. Additionally, non-conducting dielectric cell containing atomic samples reduces the field perturbation comparing with metal sensing head used in traditional probe [24,35-37]. Based on the reasons mentioned above, atom-based measurement provides possibility of detecting the MW field with higher accuracy.

In this paper, we present an alternate technique for measuring magnetic field inside MW cavity, using atomic Rabi resonances. In our technique, a phase-modulated MW field drives atomic Rabi resonance, which is a function of the modulation frequency $\omega_m$ and displays its peak when $\omega_m = \Omega/2$. By scanning $\omega_m$ for a given input power, we can obtain the resonance lineshape, then $\Omega$ (i.e., field strength) is determined by fitting the measured data to the theoretical model of field measurement (see part II) [38]. We performed a series of experiments with the proposed atom-based field detection technique to measure a certain range of MW power at about 9.2 GHz using our cavity-based Cs Rabi resonances experimental setup. The agreement of the experimental results and theoretical calculations proves our technique is simple and effective. The technique provides a direct means for accurate measurement of magnetic field strength inside cell in the gas-type atomic clocks, where traditional techniques are unavailable due to the presence of vapor cell [31,32,39]. We note also that MW magnetic field can be measured by means of Rabi oscillations [29-34]. Rabi oscillations exhibit multi-peak feature in the measurement, in which the Rabi frequency for a given power level is determined by fitting the data to a cosine function. As we will show in detail below, field detection model based on Rabi Resonance is a single-peak function, making it easy to perform field measurement. Additionally, we perform this experiment with Cs atoms, because of many practical requirements for MW measurements often occurring in X-band.

## II. FIELD MEASUREMENT THEORY

The Rabi resonances were intensely studied by Camparo et al., including the $\alpha$ and $\beta$ Rabi resonances [40,41]. The basic principle of Rabi resonances theory is given here, more details on the theory are offered in Ref. [41]. When two-level atoms interact with a phase-modulated MW field, the probability of finding the atoms in the excited-state oscillates at the modulation frequency, $\omega_m$, and its second harmonic, $2\omega_m$. The amplitudes of this oscillating probability are resonant functions of field strength, and display resonant enhancements when $\Omega = \omega_m$ and $\Omega = 2\omega_m$, these resonant enhancements of oscillation amplitudes are termed the α Rabi resonance and the β Rabi resonance, respectively. Under small-signal assumption, the oscillation amplitudes of the two Rabi resonances are given by the followings [41]:

$$P_\alpha^0 = -\frac{1}{2} \frac{m\omega_m \Omega^2 \Delta}{\left[\gamma_2^2 + \Delta^2 + \left(\frac{\gamma_2}{\gamma_1}\right)\Omega^2\right]\sqrt{\left(\Omega^2 - \omega_m^2\right)^2 + \gamma_1^2 \omega_m^2}}, \quad (1a)$$

$$P_\beta^0 = \frac{1}{4} \frac{m^2 \omega_m \Omega^2 \gamma_2}{\left[\gamma_2^2 + \Delta^2 + \left(\frac{\gamma_2}{\gamma_1}\right)\Omega^2\right]\sqrt{\left(\Omega^2 - 4\omega_m^2\right)^2 + 4\gamma_1^2 \omega_m^2}}, \quad (1b)$$

---

*Corresponding author: fysun@uestc.edu.cn

where $\gamma_1$ and $\gamma_2$ are the longitudinal and transverse relaxation rate, respectively, $\Delta$ is the average MW field-atom detuning, m is the modulation index. Hyperfine transition Rabi frequency $\Omega$ is proportional to field strength. For the magnetic dipole transition considered here, $\Omega = (\mu_B B)/\hbar$, where $\mu_B$ is Bohr magneton, $\hbar = h/2\pi$, with h being Plank's constant, and $B$ represents the MW magnetic field strength which is parallel to quantization axis of the Zeeman sublevels.

By using the Rabi resonances, several typical applications based on the theory have been proposed and/or demonstrated experimentally, including the strength stabilization of microwave field and/or laser field (so called atomic candle) [42], the measurements of material properties (e.g. absorption/refractive-index) [43], the observations of cavity mode stability for Rb atomic clock [44], and the generation of MW power standard [16-19]. In these cases, the β Rabi resonance amplitude is considered as a function of Rabi frequency where, modulation frequency is fixed during operation [41,42].

For field measurement application demonstrated here, we treat the β Rabi resonance as a function of modulation frequency and use it to measure an unknown magnetic field strength [38,41]. According to Eq. (1b), $\frac{m^2\Omega^2\gamma_2}{[\gamma_2^2+\Delta^2+(\gamma_2/\gamma_1)\Omega^2]}$ can be considered to be constant under the fixed experiment settings, including laser and MW levels. Thus, the amplitude of the β resonance has following simple expression:

$$P_\beta^0(\omega_m) \propto \frac{\omega_m}{\sqrt{(\Omega^2-4\omega_m^2)^2+4\gamma_1^2\omega_m^2}}. \quad (2)$$

From Eq. (2), the amplitude of the oscillating function, $P_\beta^0$, exhibits resonance when $\omega_m = \Omega/2$ with a peak amplitude of $1/(2\gamma_1)$. To measure unknown field strength, we first experimentally measure the resonance lineshape by scanning $\omega_m$, and then $\Omega$ can be determined by fitting theoretical model obtained above to the measured lineshape.

### III. EXPERIMENT SETUP

Our experimental setup is shown in Fig. 1(a). We use a Cs vapor cell to perform Rabi resonances between atomic hyperfine states ($|F=3, m_F=0\rangle$—$|F=4, m_F=0\rangle$). A linearly polarized $4-4'$ laser is used to pump and detect Cs atoms, whose frequency is stabilized to saturated absorption resonances on the D2 line. For all measurements presented here a laser power of $\sim 30\ \mu W$ corresponding to density of $\sim 150\ \mu W/cm^2$ is used. Using braided windings wrapped around the cavity, the temperature of the vapor cell (filled with pure Cs and 10 Torr $N_2$) is heated and stabilized at $\sim 36\ °C$. Whole physical package is centrally mounted inside three-dimensional Helmholtz coils.

We note that the detection of Rabi resonances only allows for determination of the local Rabi frequency, and hence the local MW field strength. Therefore, uniform field distribution is needed for the determination of unique Rabi frequency. To address this problem, the cell is placed in the center of rectangular cavity symmetrically about a magnetic field maximum. The inside dimensions of the rectangular cavity are $22.86\ mm \times 10.16\ mm \times 72\ mm$, which allow for the $TE_{104}$ mode operating near 9.2 GHz. The advantage of this scheme is that the MW magnetic field is approximately uniform over the entire interaction region in the atomic samples. To demonstrate this advantage, we simulate the magnetic field inside the cavity to show how the fields keep uniform in the center of the cavity, as shown in Fig. 1(b). MW-couplings are created with two small holes with diameter of $\sim 6mm$. The two waveguide to coaxial adapters are connected to the cavity. After passing through the first adapter, the input MW is irradiated into the cavity via a coupling hole to drive Rabi resonances in a thermal Cs atomic vapor. The output port of the cavity is connected to the second adapter, which is terminated by a 50 Ohm matched load. Two cutoff waveguides are attached to the cavity walls with the purpose of suppressing microwave leakage and field perturbation during operation. Through the cutoff waveguides, the laser illuminates the cell perpendicular to the direction of the applied MW magnetic field. Transmission of the laser through the vapor is monitored with a $1cm^2$ Si photodiode, and then finally recorded with a FFT Analyzer.

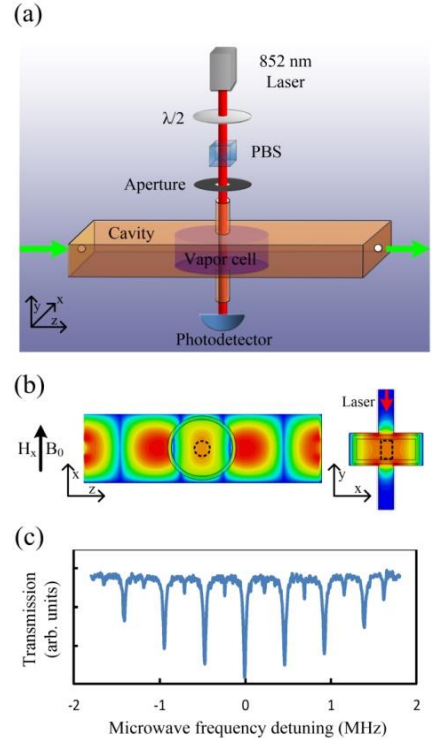

Fig. 1. (Color online) (a) Schematic of the experimental setup. The cell with a height of $\sim 10\ mm$, and a diameter of $\sim 20\ mm$, is located in the position of the maximum magnetic field. Laser from a distributed feedback (DFB) diode is tuned to the Cs D2 transition at 852 nm, and then expanded and apertured to a final diameter of $\sim 5\ mm$. Adapters, matched load, three-dimensional Helmoltz coils and heating coils are not shown here. (b) Simulated magnetic field distribution, $H_X$, along quantization axis, $B_0$, in the cavity. Left: Top view (x-z plane). Right: side view (x-y plane). MW-Atom-Laser interaction region is surrounded by black dot line. (c) Detected Double-resonance spectrum by sweeping MW frequency in the absence of phase modulation.

To validate our experimental setup, the detection of the optical-microwave double-resonance (DR) signal is the first step before field measurement [45]. We assume here that a static magnetic field of 700 mG is in the x direction, also in this direction with MW magnetic field. Under these conditions, the seven $\Delta m_F = 0$ magnetic dipole transitions are observed as frequency is scanned, and these DR spectrum are recorded via the transmission light through the cell, as illustrated in Fig. 2(c).

## IV. RESULTS

In order to measure Rabi frequency accurately, two main works need to be done before performing the actual measurement: to show how to obtain oscillation amplitudes ($P_\alpha^0$ and $P_\beta^0$); and to determine atomic transition frequency.

For the former, the obtainments of $P_\alpha^0$ and $P_\beta^0$ are direct via FFT spectrum of the Rabi resonances in the presence of the phase-modulated field. Fig. 2 shows an example of $P_\alpha^0$ and $P_\beta^0$ resonating at 10 kHz and 20 kHz, respectively, with different MW frequencies, where a 10 kHz phase modulation signal is applied to the MW field. Specifically, Fig. 2(a) shows on-resonance state ($P_\alpha^0 = 0$ when $\Delta = 0$) on which the oscillation amplitude only depends on the β Rabi resonance. From Fig, 2(b) and 2(c), which represent the near-resonance cases, we find that a slight frequency detuning could also induce MW-atoms interaction. As expected, no field-atoms interaction happens at MW frequency far from resonance (see Fig. 2(d)).

For the latter, measure of atomic transition frequency is needed because the Rabi frequency is also a function of the detuning from the resonance. Here, we use the α Rabi resonance to determine transition frequency (see Eq. (1a)) [41,42]. Prior to the excitation by the modulated field, we carry out the typical DR operation (see Fig. 1(c)) and use obtained DR spectrum to estimate the transition frequency. After that, the function of phase modulation is turned on, then we tune MW frequency finely (around the estimation frequency) till $P_\alpha^0 = 0$, while keeping $P_\beta^0 \neq 0$ during the tuning process. And obviously, the atomic transition frequency is slightly shifted from 9.19263177 GHz due to the presence of $N_2$. Thus, MW frequency which is required for interaction is precisely confirmed through the procedures described above.

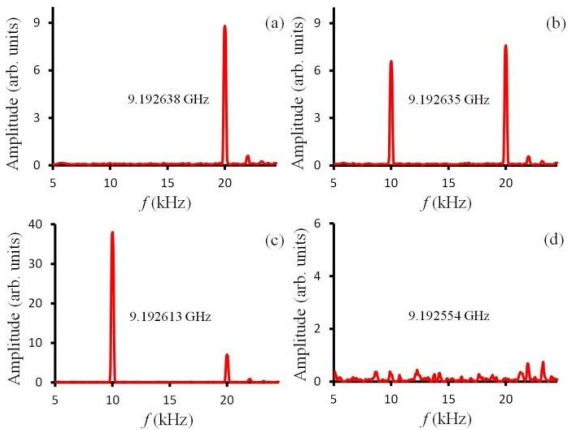

Fig. 2. (Color online) Fourier spectrum of the transmitted laser intensity in the presence of phase modulation under different MW frequencies. (a) On-resonance. $P_\alpha^0 = 0$, $P_\beta^0 \neq 0$. (b) and (c) Near-resonance. $P_\alpha^0 \neq 0$, $P_\beta^0 \neq 0$. (d) Non-resonance. $P_\alpha^0 = 0$, $P_\beta^0 = 0$.

Next, to validate our theoretical model, we will show how the Rabi frequency can be determined by this technique. As an illustrative example, Fig. 3 presents a comparison of theoretical and experimentally measured lineshapes of the β Rabi resonances at two different power levels. The experimental data are obtained by measuring $P_\beta^0$ oscillations occurring at $2\omega_m$ on the FFT spectrum analyzer as a function of $\omega_m$. The solid curves in this figure represent the theoretical fit to the experimental data using Eq. (2), with their peaks yield information about the Rabi frequencies. As can be seen from Fig. 3, the measured data closely follow the theory curves.

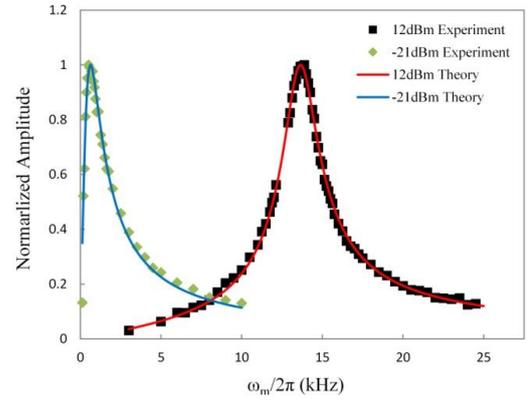

Fig. 3. (Color online) Typical Rabi resonance signals for measuring Rabi frequency. Experimental points (dots) and theoretical fit (solid line) are shown. The theory is fit to the experimental data by using Eq. (2) and varying the vertical scale.

For completeness and quantitative understanding of our proposed method, Fig. 4 presents the power-dependent Rabi resonance signals as a function of modulation frequency at six different power levels of P = -12 dBm, -6 dBm, 0 dBm, 6 dBm, 12 dBm, and 18 dBm without normalizing the resonance amplitudes, which correspond to Rabi frequencies of $\Omega / 2\pi$ = 2.06 kHz, 3.54 kHz, 6.79 kHz, 13.49 kHz, 27.3 kHz, and 54.46 kHz, respectively. From Fig. 4, we find that for every 6 dB increases in P, the Rabi frequency will be approximately doubled. This relationship is valid especially for a relative large power input. It is worth to note that the validity of the small-signal assumption ($\Omega \gg \gamma_1$) is gradually weakened as the applied power decreases, and this causes the decline of measurement accuracy. Despite this, our method still exhibits a reasonable approximation. Additionally, signal amplitude varies rapidly with the variation of applied power. However, what we concern most is the frequency corresponding to resonance peak rather than the amplitude in these measurements. This a major advantage of atom-based measurement. Clearly, Rabi resonance lineshape has a unique peak at $\omega_m = \Omega / 2$ for any given $\Omega$, as predicted by Eq. (2).

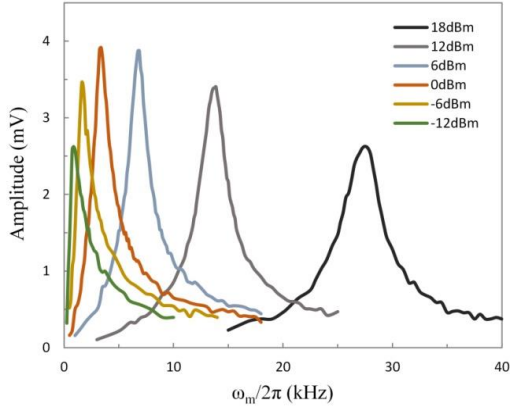

Fig. 4. (Color online) Rabi resonance signals versus modulation frequencies with different applied power levels at 9.2 GHz.

A more detailed analysis of lineshapes from Fig. 3 and Fig. 4 shows that Rabi resonance lineshape has a certain width. Here, similar to Q definition for the atomic candle,[41] we let $Q = \Omega/2\Delta\omega_{FWHM}$ for field measurement model (Eq. (2)), where $\Delta\omega_{FWHM}$ is the full width at half maximum (FWHM). Knowing that the resonance peak occurs at $\omega_m = \Omega/2$ with an amplitude of $1/(2\gamma_1)$, we set $P_\beta^0[(\Omega-\Delta\omega_{FWHM})/2] = 1/(4\gamma_1)$. Then, according to Eq. (2), $\Delta\omega_{FWHM}$ can be computed as

$$\Delta\omega_{FWHM} = \frac{\sqrt{2\Omega^2 + 3\gamma_1^2 + \gamma_1\sqrt{12\Omega^2 + 9\gamma_1^2}} - \sqrt{2\Omega^2 + 3\gamma_1^2 - \gamma_1\sqrt{12\Omega^2 + 9\gamma_1^2}}}{2\sqrt{2}}, \quad (3)$$

Thus, the line Q has the potential to be narrowed greatly by decreasing the longitudinal relaxation rate.

Following the technique described above, in Fig. 5, we repeat the measurements while varying the incident power level, P in the range from -21 dBm (7.94 μW) to 20 dBm (100 mW). We then plot the Rabi frequencies as a function of power and fit the linear function of our measurements. Since the magnetic field sensed by atoms is proportional to the square root of the incident MW power, $\sqrt{P}$, in case of cavity structure, thus, as one would expect, the measured Rabi frequency shows linear relationship with the square root of the incident MW power. From Fig. 5, it is clear that our setup serves as a field/power sensor in these measurements.

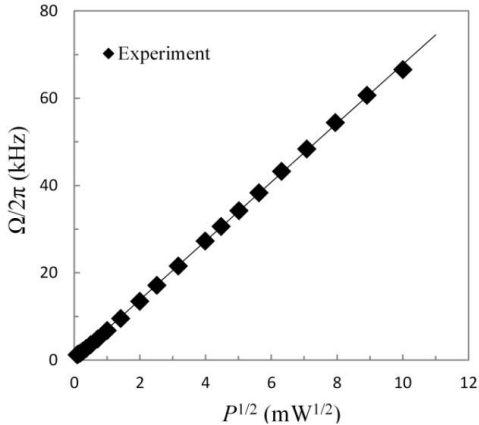

Fig. 5. Rabi frequencies as a function of applied MW power. The black solid line is linear fit to the measured values.

## V. SIMULATION AND COMPARISON

To further verify the validity of this measurement technique, we use HFSS (finite element simulation tool) to calculate the field strength inside the interaction region, and make a comparison to measured results. It is challenging to determine the wall thickness and the dielectric characteristic of the vapor cell accurately due to manufacturing limitation [21,22,25]. In our simulation, we adjust the two parameters slightly to match real measurement conditions which mainly including the measurement frequency and cavity resonant frequency. This is particular important for accurate simulation, since field distribution is frequency-dependent due to resonant character of the cavity. Fig. 6 shows simulated MW magnetic field along three directions of the cavity with a power injection of 20 dBm, indicating that the field distribution is quite uniform over entire interaction region (i.e., -2.5 mm≤ x,z ≤ 2.5mm, and -3.3 mm≤ y ≤ 3.3mm), where the center of the cell is defined as the origin of coordinates (0, 0, 0). Only a field variation of ~ 7.5 % is observed in y-direction but there are almost no variations in x- and z-directions due to the narrow laser diameter (~ 5 mm). A significant field perturbation is observed due to the presence of the cell and optical holes. Fig. 6 also indicates how to determine simulated field. Fig. 7 shows a comparison between the simulation results and those obtained from the practical measurements at different power levels, where experimental values are calculated from Rabi frequencies shown in Fig. 5. From Fig. 7, similar linear behaviors are clearly observed from both methods as expected.

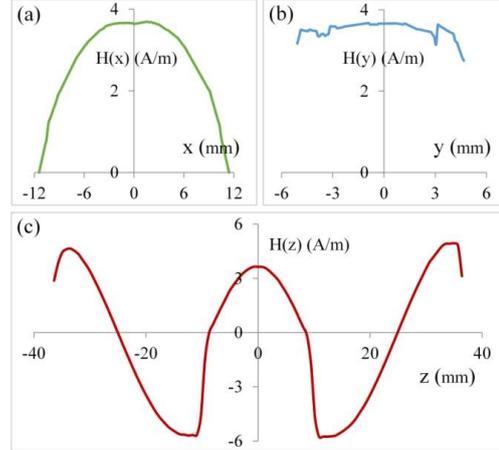

Fig. 6. (Color online) Simulated magnetic field strength along the quantization axis as a function of position. (a) Field strength along x direction (y=z=0). (b) Field strength along y direction (x=z=0). (c) Field strength along z direction (x=y=0), clearly, the cavity is operated in TE$_{104}$ mode.

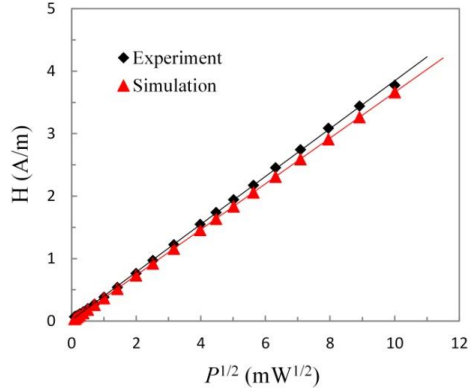

Fig. 7. (Color online) Comparison of experimental and simulated magnetic field strength as a function of applied MW power. The black solid line and the red solid line are linear fits to the measured values and simulated values, respectively.

## VI. DISCUSSION AND CONCLUSION

Note that no attempt has been made to optimize this setup to obtain ultimate ability to detect MW field. Clearly that this ability can be easily extended to cover not only broader scope of under-test field/power, but also other frequency bands by using different atomic species and/or energy level transitions (especially for the use of magnetic-sensitive hyperfine transition). It is worth noting that the technique proposed here is demonstrated how to measure field strength in a MW cavity via Rabi resonances. In fact, this technique is also available for near/far-field cases. From a practical point of view, the continuous interaction of atoms with both MW and laser makes Rabi resonances measurement easy to implement compared to traditional Rabi oscillations measurement, in which MW and laser are pulse-modulated. This offers the possibility of using this technique for the measure of MW magnetic field without the requirement of complicated system.

In summary, we demonstrate a technique for measuring MW field inside cavity via atomic dynamical Rabi resonances. To test the validity of this technique, we perform a demonstrating experiment, and we compare the power levels measured from the experiments to those computed from the theoretical model. Both theoretical and experimental results demonstrate that, in principle, Rabi resonance can be used to measure field strength. Finally, we emphasize that the Rabi resonances technique, due to its simplicity and single-peak character, may offer a very useful technique to measure MW field strength. Using this atom-based, SI-traceable measurement technique, MW power can be linked directly to the modulation frequency, which could be referenced to a primary frequency standard.

## References

[1] J. Kitching, S. Knappe, and E. A. Donley, atomic sensors-a review, IEEE J. Sens. **11**, 1749 (2011).

[2] D. Budker and M. Romalis, optical magnetometry, Nat. Phys. **3**, 227 (2007).

[3] K. Cox, V. I. Yudin, A. V. Taichenachev, I. Novikova, and E. E. Mikhailov, Measurements of the magnetic field vector using multiple electromagnetically induced transparency resonances in Rb vapor, Phys. Rev. A **83**, 015801 (2011).

[4] R. Mhaskar, S. Knappe, and J. Kitching, A low-power, high-sensitivity micromachined optical magnetometer, Appl. Phys. Lett. **101**, 241105 (2012).

[5] D. Sheng, S. Li, N. Dural, and M. V. Romalis, Subfemtotesla Scalar Atomic Magnetometry Using Multipass Cells, Phys. Rev. Lett. **110**, 160802 (2013).

[6] A. Osterwalder, and F. Merkt, Using High Rydberg States as Electric Field Sensors, Phys. Rev. Lett. **82**, 1831 (1999).

[7] A. Facon, E. K. Dietsche. D. Grosso, S. Haroche, J. M. Raimond, M. Brune, and S. Gleyzes, A sensitive electrometer based on a Rydberg atom in a Schrödinger-cat state, Nature **535**, 262 (2016).

[8] I. M. Savukov, S. J. Seltzer, M. V. Romalis, and K. L. Sauer, Tunable Atomic Magnetometer for Detection of Radio-Frequency Magnetic Fields, Phys. Rev. Lett. **95**, 063004 (2005).

[9] S. K. Lee, K. L. Sauer, S. J. Seltzer, O. Alem, and M. V. Romalis, Subfemtotesla radio-frequency atomic magnetometer of nuclear quadrupole resonanceAppl. Phys. Lett. **89**, 214106 (2006).

[10] G. Katsoprinakis, D. Petrosyan, and I. K. Kominis, High Frequency Atomic Magnetometer by Use of Electromagnetically Induced Transparency, Phys. Rev. Lett. **97**, 230801 (2006).

[11] M. P. Ledbetter, V. M. Acosta, S. M. Rochester, D. Budker, S. Pustelny, and V. V. Yashchuk, Detection of radio-frequency magnetic fields using nonlinear magneto-optical rotation, Phys. Rev. A **75**, 023405 (2007).

[12] W. Wasilewski, K. Jensen, H. Krauter, J. J. Renema, M. V. Balabas, and E. S. Polzik, Quantum Noise Limited and Entanglement-Assisted Magnetometry, Phys. Rev. Lett. **104**, 133601 (2010).

[13] W. Chalupczak, R. M. Godun, S. Pustelny, and W. Gawlik, Room temperature femtotesla radio-frequency atomic magnetometer, Appl. Phys. Lett. **100**, 242401 (2012).

[14] T. P. Crowley, E. A. Donley, and T.P. Heavner, Quantum-based microwave power measurements: Proof-of-concept experiment, Rev. Sci. Instrum. **75**, 2575 (2004).

[15] D. C. Paulusse, N. L. Rowell, and A. Michaud, Accuracy of an Atomic Microwave Power Standard, IEEE Trans. Instrum. Meas. **54**, 692 (2005)

[16] M. Kinoshita, K. Shimaoka, and K. Komiyama, Determination of the Microwave Field Strength Using the Rabi Oscillation for a New Microwave Power Standard.IEEE Trans. Instrum. Meas. **58**, 1114 (2009)

[17] M. Kinoshita, K. Shimaoka, and K. Komiyama, Atomic Microwave Power Standard Based on the Rabi Frequency, IEEE Trans. Instrum. Meas. **60**, 2696 (2011)

[18] M. Kinoshita, K. Shimaoka, and Y. Shimada, Optimization of the Atomic Candle Signal for the Precise Measurement of Microwave Power, IEEE Trans. Instrum. Meas. **62**, 1807(2013)

[19] J. A. Sedlacek, A. Schwettmann, H. Kübler, R. Löw, T. Pfau, and J. P. Shaffer, Microwave electrometry with Rydberg atoms in a vapour cell using bright atomic resonances, Nat. Phys. **8**, 819 (2012).

[20] J. A. Sedlacek, A. Schwettmann, H. Kübler, and J. P. Shaffer, Atom-Based Vector Microwave Electrometry Using Rubidium Rydberg Atoms in a Vapor Cell, Phys. Rev. Lett. **111**, 063001 (2013).

[21] C. L. Holloway, J. A. Gordon, A. Schwarzkopf, D. A. Anderson, S. A. Miller, N. Thaicharoen, and G. Raithel, Sub-wavelength imaging and field mapping via electromagnetically induced transparency and Autler-Townes splitting in Rydberg atoms, Appl. Phys. Lett. **104**, 244102 (2014).

[22] J. A. Gordon, C. L. Holloway, A. Schwarzkopf, D. A. Anderson, S. Miller, N. Thaicharoen, and G. Raithel, Millimeter wave detection via Autler-Townes splitting in rubidium Rydberg atoms, Appl. Phys. Lett. **105**, 024104 (2014);

[23] C. L. Holloway, J. A. Gordon, S. Jefferts, A. Schwarzkopf, D. A. Anderson, S. A. Miller, N. Thaicharoen, and G. Raithel, Broadband Rydberg Atom-Based Electric-Field Probe for SI-Traceable, Self-Calibrated Measurements, IEEE Trans. Antennas Propag. **62**, 6169 (2014).

[24] H. Q. Fan, S. Kumar, J. Sedlacek, H. Kübler, S. Karimkashi, and J. P. Shaffer, Atom based RF electric field sensing, J. Phys. B **48**, 202001 (2015).

[25] H. Q. Fan, S. Kumar, J. T. Sheng, J. P. Shaffer, C. L. Holloway, and J. A. Gordon, Effect of Vapor-Cell Geometry on Rydberg-Atom-Based Measurements, Phys. Rev.

Appl. **4**, 044015 (2015).

[26] D. A. Anderson, S. A. Miller, and G. Raithel, J. A. Gordon, M. L. Butler, and C. L. Holloway, Optical Measurements of Strong Microwave Fields with Rydberg Atoms in a Vapor Cell, Phys. Rev. Appl. **5**, 034003 (2016);

[27] M. T. Simons, J. A. Gordon, C. L. Holloway, D. A. Anderson, S. A. Miller, and G. Raithel, Using frequency detuning to improve the sensitivity of electric field electromagnetically induced transparency and Autler-Townes splitting in Rydberg atoms, Appl. Phys. Lett. **108**, 174101 (2016).

[28] H. Q. Fan, S. Kumar, R. Daschner, H. Kübler, and J. P. Shaffer, Subwavelength microwave electric-field imaging using Rydberg atoms inside atomic vapor cells, Opt. Lett. **39**, 3030 (2014).

[29] P. Böhi, M. F. Riedel, T. W. Hänsch, and P. Treutlein, Imaging of microwave fields using ultracold atoms, Appl. Phys. Lett. **97**, 051101 (2010).

[30] P. Böhi and P. Treutlein, Simple microwave field imaging technique using hot atomic vapor cells, Appl. Phys. Lett. **101**, 181107 (2012).

[31] A. Horsley, G. X. Du, M. Pellaton, C. Affolderbach, G. Mileti, and P. Treutlein, Imaging of relaxation times and microwave field strength in a microfabricated vapor cell, Phys. Rev. A **88**, 063407 (2013).

[32] C. Affolderbach, G. X. Du, T. Bandi, A. Horsley, P. Treutlein, and G. Mileti, Imaging Microwave and DC Magnetic Fields in a Vapor-Cell Rb Atomic Clock IEEE Trans. Instrum. Meas. **64**, 3629 (2015).

[33] A. Horsley, G.X. Du, and P. Treutlein, Widefield microwave imaging in alkali vapor cells with sub-100μm resolution, New J. Phys. **17**, 112002 (2015).

[34] A. Horsley and P. Treutlein, Frequency-tunable microwave field detection in an atomic vapor cell, Appl. Phys. Lett. **108**, 211102 (2016).

[35] N. S. Nahman, M, Kanda, E. B. Larsen, and M. L. Crawford, Methodology of measurement of standard electricitty and magnetic field, IEEE Trans. Instrum. Meas. **IM-34**, 490 (1985).

[36] B. Mellouet, L. Velasco, and J. Achkar, Fast Method Applied to the Measurement of Microwave Power Standards, IEEE Trans. Instrum. Meas. **50**, 381 (2001).

[37] A. S. Brush, Measurement of microwave power-A review of techniques used for measurement of high-frequency RF power, IEEE Instrum. Meas. Mag. **10**, 20 (2007)

[38] M. Kinoshita, K. Shimaoka, and K. Komiyama, Rabi frequency measurement for microwave power standard using double resonance spectrum in Proc. CPEM Dig., Jun. 2008, pp. 698–699.

[39] R. P. Frueholz, and J. C. Camparo, Microwave field strength measurement in a rubidium clock cavity via adiabatic rapid, J. Appl. Phys. **57**, 704 (1985).

[40] J. C. Camparo and R. P. Frueholz, Observation of the Rabi-resonance spectrum, Phys. Rev. A **38**, 6143 (1988).

[41] J. G. Coffer, B. Sickmiller, A. Presser, and J. C. Camparo, Line shapes of atomic-candle-type Rabi resonances, Phys. Rev. A **66**, 023806 (2002).

[42] J. C. Camparo, Atomic Stabilization of Electromagnetic Field Strength Using Rabi Resonances, Phys. Rev. Lett. **80**, 222 (1998).

[43] T. S. Wood, J. G. Coffer, and J. C. Camparo, Precision Measurements of Absorption and Refractive-Index Using an Atomic Candle, IEEE Trans. Instrum. Meas. **50**, 1229 (2001).

[44] J. G. Coffer, B. Sickmiller, and J. C. Camparo, Cavity-Q Aging Observed via an Atomic-Candle Signal, IEEE Trans. Ultrason. Ferroelectr. Freq. Control. **51**, 139 (2004).

[45] M. Pellaton, C. Affolderbach, Y. Pétremand, N. de Rooij, and G. Mileti, Study of laser-pumped double-resonance clock signals using a microfaricated cell, Phys. Scr. **T149**, 014013 (2012).